\documentclass{LMCS}
\usepackage[T1]{fontenc}
\usepackage[latin1]{inputenc}
\usepackage{subfigure}
\usepackage{graphicx}
\usepackage{amssymb}

\makeatletter

\providecommand{\tabularnewline}{\\}

  \theoremstyle{plain}
  \newtheorem{fAct}[thm]{fAct}
  \theoremstyle{definition}
  \newtheorem{defn}[thm]{Definition}
  \theoremstyle{remark}
  \newtheorem*{acknowledgement*}{Acknowledgement}

\usepackage{subfigure}
\usepackage{hyperref}

\def\buchi{B\"{u}chi}
\DeclareMathOperator{\NFW}{NFW}

\DeclareMathOperator{\NBW}{NBW}
\DeclareMathOperator{\NGBW}{NGBW}
\DeclareMathOperator{\NSW}{NSW}

\DeclareMathOperator{\DRW}{DRW}
\DeclareMathOperator{\DSW}{DSW}
\def\dprod{\mathop{\displaystyle \prod }}%
\def\QATOPD#1#2#3#4{{#3 \atopwithdelims#1#2 #4}}

\def\doi{4 (1:?) 2008}
\lmcsheading%
{\doi}
{1--20}
{}
{}
{Jul.~25, 2007}
{Mar.~??, 2008}
{}   

\begin{document}

\title[Lower Bounds for Complementation of $\omega\,$-Automata]{Lower Bounds for Complementation of $\omega\,$-Automata via the Full
Automata Technique\rsuper *}

\author[Q.~Yan]{Qiqi Yan}

\address{Department of Computer Science and Engineering, Shanghai Jiao Tong
University, 200240, Shanghai, P.R. China}

\email{contact@qiqiyan.com}
\thanks{Supported by NSFC No. 60273050.}

\titlecomment{{\lsuper *}A preliminary version of this paper appears
  in the proceedings of the 33rd International Colloquium on Automata, 
  Languages and Programming, 2006.}

\begin{abstract}
In this paper, we first introduce a lower bound technique for the
state complexity of transformations of automata. Namely we suggest
first considering the class of full automata in lower bound analysis,
and later reducing the size of the large alphabet via alphabet substitutions.
Then we apply such technique to the complementation of nondeterministic
$\omega$-automata, and obtain several lower bound results. Particularly,
we prove an $\Omega((0.76n)^{n})$ lower bound for  \buchi\ complementation,
which also holds for almost every complementation or determinization
transformation of nondeterministic $\omega$-automata, and prove an
optimal $(\Omega(nk))^{n}$ lower bound for the complementation of
generalized \buchi\ automata, which holds for Streett automata as
well. 
 
\end{abstract}
\maketitle

\section{Introduction\label{sec:intro}}

The complementation problem of nondeterministic $\omega$-automata,
i.e.\ nondeterministic automata over infinite words, has various
applications in formal verification. For example in automata-theoretic
model checking, in order to check whether a system represented by
automaton $\mathcal{A}_{1}$ satisfies a property represented by automaton
$\mathcal{A}_{2}$, one checks that the intersection of $\mathcal{A}_{1}$
with an automaton that complements $\mathcal{A}_{2}$ is an automaton
accepting the empty language \cite{Kur94,VW94}. In such a process,
several types of nondeterministic $\omega$-automata are concerned,
including \buchi, generalized \buchi, Rabin, Streett etc., and the
complexity of complementing these automata has caught great attention.

The complementation of \buchi\ automata has been investigated for
over forty years \cite{Var07}. The first effective construction was
given in \cite{Buc62}, and the first exponential construction was
given in \cite{SVW85} with a $2^{O(n^{2})}$ state blow-up ($n$
is the number of states of the input automaton). Even better constructions
with $2^{O(n\log n)}$ state blow-ups were given in \cite{Saf88,Kla91,KV01},
which match with Michel's $n!=2^{\Omega(n\log n)}$ lower bound \cite{Mic88},
and were thus considered optimal. However, a closer look reveals that
the blow-up of the construction in \cite{KV01} is $(6n)^{n}$, while
Michel's lower bound is only roughly $(n/e)^{n}=(0.36n)^{n}$, leaving
a big exponential gap hiding in the asymptotic notation%
\footnote{In contrast, for the complementation of nondeterministic finite automata
over finite words, the $2^{n}$ blow-up of the subset construction
\cite{RS59} was justified by a tight lower bound \cite{SS78}, which
works even if the alphabet concerned is binary \cite{Jir05}.%
}. Motivated by this complexity gap, the construction in \cite{KV01}
was further refined in \cite{FKV06} to $(0.97n)^{n}$. On the other
hand, Michel's lower bound was never improved.

For generalized \buchi, Rabin and Streett automata, the best known
constructions are in \cite{KV05,KV05a}, which are $2^{O(n\log nk)}$,
$2^{O(nk\log n)}$ and $2^{O(nk\log nk)}$ respectively. Here state
blow-ups are measured in terms of both $n$ and $k$, where $k$ is
the index of the input automaton. Optimality problems of these constructions
have been vastly open, because only $2^{\Omega(n\log n)}$ lower bounds
were known by variants of Michel's proof \cite{Lod99}.

What remains missing are stronger lower bound results. Tighter lower
bounds usually lead us into better understanding of the intricacy
of the complementation of nondeterministic $\omega$-automata, and
are the main concern of this paper. Such understanding can suggest
methods to further optimize the constructions, or to circumvent those
difficult cases in practice.

\begin{figure}[h]
\centering\includegraphics[scale=1.3]{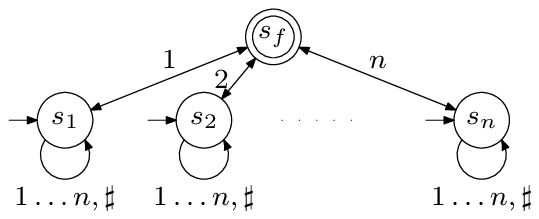}

\caption{\label{fig:michel}Michel's Automata Class}
\end{figure}

To understand why we have so few strong lower bounds, we observe that
at the core of almost every known lower bound is Michel's result,
which was obtained in the traditional way. That is, one first constructs
a particular family of automata $(\mathcal{A}_{n})_{n\geq1}$, and
then proves that complementing each $\mathcal{A}_{n}$ requires a
large state blow-up. The $\mathcal{A}_{n+1}$ of Michel's automata
family is depicted in Figure \ref{fig:michel}. Although each $\mathcal{A}_{n+1}$
has a simple structure, it is not straightforward to see what language
it accepts, and nor is it clear at all how we can work with this automaton
for lower bound. %

In many cases, identifying such an automata family is difficult, and
is the main obstacle towards lower bounds. In this paper, we propose
a new technique to circumvent this difficulty. Namely, we suggest
first considering the family of full automata in lower bound analysis,
and later reducing the size of the large alphabet via alphabet substitutions.
A simple demonstration of such technique is presented in Section \ref{sec:full}.

With the help of full automata, we tighten the state complexity $BC(n)$
of \buchi\ complementation from $(0.36n)^{n}\leq BC(n)\leq(0.97n)^{n}$
to $(0.76n)^{n}\leq BC(n)\leq(0.97n)^{n}$. Surprisingly, this $(0.76n)^{n}$
lower bound also holds for every complementation or determinization
transformation concerning \buchi, generalized \buchi, Rabin, Streett,
Muller, and parity automata. As to the complementation of generalized
\buchi\ automata, we prove an $(\Omega(nk))^{n}$ lower bound, matching
with the $(O(nk))^{n}$ upper bound in \cite{KV05}. This lower bound
also holds for the complementation of Streett automata and the determinization
of generalized \buchi\ automata into Rabin automata. A summary of
our lower bounds is given in Section \ref{sec:summary}.

\subsection*{Full Automata and Sakoda and Sipser's Languages}

It turns out that the notion of full automata is similar to Sakoda
and Sipser's languages in \cite{SS78}. Their language $\mathcal{B}_{n}$
actually corresponds to the $\Delta$-graphs of the words accepted
by some full automaton. Also as pointed to us by Christos A. Kapoutsis,
the technique of alphabet substitution was somewhat implicit in Sakoda
and Sipser's paper (but presented in a somewhat obscure way, refer
to the paragraph before their Theorem 4.3.2). So the full automata
technique is more like a new treatment of some techniques in the Sakoda
and Sipser's paper, rather than a totally new invention. Compared
to Sakoda and Sipser's languages, the notion of full automata enjoys
a simple definition and is very handy to use. It is also more readily
to be extended to other kinds of automata like alternating automata.

For unclear reasons, Sakoda and Sipser's languages were rarely applied
to fields other than 2-way automata after their paper. We hope that
our treatment will make a clear exposition of the techniques and demonstrate
their usefulness in problems on automata over one-way inputs as well.

\section{Basic Definitions}\label{sec:defs}

A (\emph{nondeterministic}) \emph{automaton} is a tuple $\mathcal{A}=(\Sigma,S,I,\Delta,\ast)$
with alphabet $\Sigma$, finite state set $S$, initial state set
$I\subseteq S$, transition relation $\Delta\subseteq S\times\Sigma\times S$
and $\ast$ some extra components. Particularly $\mathcal{A}$ is
\emph{deterministic} if $\vert I\vert=1$ and for all $p\in S$ and
$a\in\Sigma$, $\vert\{ q\in S\mid\langle p,a,q\rangle\in\Delta\}\vert\leq1$.

For a word $w=a(0)a(1)\dots a(l-1)\in\Sigma^{\ast}$ with $length(w)=l\geq0$,
a \emph{}finite run \emph{}of $\mathcal{A}$ from state $p$ to $q$
over $w$ is a finite state sequence $\rho=\rho(0)\rho(1)\dots\rho(l)\in S^{\ast}$
such that $\rho(0)=p$, $\rho(l)=q$ and $\langle\rho(i),a(i),\rho(i+1)\rangle\in\Delta$
for all $0\leq i<l$. We say that $\rho$ \emph{visits} a state set
$T$ if $\rho(i)\in T$ for some $0\leq i\leq l$. We write $p\overset{w}{\longrightarrow}q$
if a finite run from $p$ to $q$ over $w$ exists, and $p\overset{w}{\underset{T}{\longrightarrow}}q$
if in addition the run visits $T$.

A (\emph{Nondeterministic}) \emph{Finite Word Automaton} ($\NFW$
for short) is an automaton $\mathcal{A}=(\Sigma,S,I,\Delta,F)$ with
final state set $F\subseteq S$. A finite word $w$ is accepted by
$\mathcal{A}$ if there is a finite run over $w$ from an initial
state to a final state. The language accepted by $\mathcal{A}$, denoted
by $\mathcal{\mathcal{L}}(\mathcal{\mathcal{A}})$, is the set of
words accepted by $\mathcal{A}$, and its complement $\Sigma^{\ast}\backslash\mathcal{\mathcal{L}}(\mathcal{\mathcal{A}})$
is denoted by $\mathcal{L}^{C}(\mathcal{A})$.

For an \emph{$\omega$}-word $\alpha=\alpha(0)\alpha(1)\dots\in\Sigma^{\omega}$,
i.e., an infinite sequence of letters in $\Sigma$, a (infinite) run
of $\mathcal{A}$ over $\alpha$ is an infinite state sequence $\rho=\rho(0)\rho(1)\dots\in S^{\omega}$
such that $\rho(0)\in I$ and $\langle\rho(i),\alpha(i),\rho(i+1)\rangle\in\Delta$
for all $i\geq0$. We let $Occ(\rho)=\{ q\in S\mid\rho(i)=q\text{ for some }i\in\mathbb{N}\}$,
$Inf(\rho)=\{ q\in S\mid\rho(i)=q$ for infinitely many $i\in\mathbb{N\}}$,
and write $\rho[l_{1},l_{2}]$ to denote the infix $\rho(l_{1})\rho(l_{1}+1)\dots\rho(l_{2})$
of $\rho$.

An (nondeterministic) \emph{$\omega$-automaton} is an automaton $\mathcal{A}=(\Sigma,S,I,\Delta,Acc)$
with acceptance condition $Acc$, which is used to decide if a run
$\rho$ of $\mathcal{A}$ is successful. There are many types of $\omega$-automata
considered in the literature \cite{Tho90}. Here we consider six of
the most common types:

\begin{itemize}
\item \emph{\buchi\ automaton}, where $Acc=F\subseteq S$ is a final state
set, and $\rho$ is successful if $Inf(\rho)\cap F\neq\emptyset$.
\item \emph{generalized \buchi\ automaton}, where $Acc=\{ F_{1},\dots,F_{k}\}$
is a list of final state sets, and $\rho$ is successful if $Inf(\rho)\cap F_{i}\neq\emptyset$
for all $1\leq i\leq k$.
\item \emph{Rabin automaton}, where $Acc=\{\langle G_{1},B_{1}\rangle,\dots,\langle G_{k},B_{k}\rangle\}$
is a list of pairs of state sets, and $\rho$ is successful if for
some $1\leq i\leq k$, $Inf(\rho)\cap G_{i}\neq\emptyset$ and $Inf(\rho)\cap B_{i}=\emptyset$.
\item \emph{Streett automaton}, where $Acc=\{\langle G_{1},B_{1}\rangle,\dots,\langle G_{k},B_{k}\rangle\}$
is a list of pairs of state sets, and $\rho$ is successful if for
all $1\leq i\leq k$, if $Inf(\rho)\cap B_{i}\neq\emptyset$, then
$Inf(\rho)\cap G_{i}\neq\emptyset$.
\item \emph{Muller automaton}, where $Acc=\mathcal{F}\subseteq\mathcal{P}owerset(S)$
is a set of state sets, and $\rho$ is successful if $Inf(\rho)\in\mathcal{F}$.
\item \emph{parity automaton}, where $Acc$ is a mapping $c:S\to\{0\dots l\}$,
and $\rho$ is successful if $\min\{ c(q)|q\in Inf(\rho)\}$ is even.
\end{itemize}
An $\omega$-word $\alpha$ is \emph{accepted} by $\mathcal{A}$ if
it has a successful run. The $\omega$\emph{-language} accepted by
$\mathcal{A}$, denoted by $\mathcal{L}(\mathcal{A})$, is the set
of $\omega$-words accepted by $\mathcal{A}$, and its complement$\
\Sigma^{\omega}\backslash\mathcal{L}(\mathcal{A})$ is denoted by $\mathcal{L}^{C}(\mathcal{A})$. The number $k$, if
defined, is called the \emph{index} of $\mathcal{A}$.

We refer to the above six types of $\omega$-automata as the \emph{common
types}. Following the convention in \cite{KV05a}, we will use acronyms
like $\NBW$, $\NGBW$, NRW etc.\ to refer to Nondeterministic \buchi/generalized
\buchi/Rabin/etc.\ Word automata. Two simple facts about these common
types of $\omega$-automata are useful for us:

\begin{fAct}
\label{fac:Loding}\cite{Lod99}(1) For every $\NBW$ $\mathcal{A}$
and every common type \emph{$\mathcal{T}$}, there exists an $\mathcal{T}$
automaton $\mathcal{A}'$ with the same number of states such that
$\mathcal{A}'$ is equivalent to $\mathcal{A}$. 

(2) For every deterministic $\omega$-automaton $\mathcal{A}$ of
a common type $\mathcal{T}$ which is not \buchi\emph{\ }nor generalized
\buchi, there exists a deterministic $\omega$-automaton $\mathcal{A}'$
of a common type (not necessarily also $\mathcal{T}$) with the same
number of states (and index, if applicable) such that $\mathcal{A}'$
complements $\mathcal{A}$.
\end{fAct}
To visualize the behavior of automata over input words, we introduce
the notion of $\Delta$-graphs. If $\mathcal{A}=(\Sigma,S,I,\Delta,\ast)$
is an automaton, then for a finite word $w=a(0)a(1)\dots a(l-1)\in\Sigma^{\ast}$
of length $l$, or an $\omega$-word $w=a(0)a(1)\dots\in\Sigma^{\omega}$
of length $l=\infty$, the $\Delta$\emph{-graph} of $w$ under $\mathcal{A}$
is the directed graph $\mathcal{G}_{w}^{\mathcal{A}}=(V_{w}^{\mathcal{A}},E_{w}^{\mathcal{A}})$
with vertex set $V_{w}^{\mathcal{A}}=\{\langle p,i\rangle\mid p\in S,0\leq i\leq l,i\in\mathbb{N}\}$
and edge set $E_{w}^{\mathcal{A}}$ defined as: for all $p,q\in S$
and $0\leq i<l$, $\langle\langle p,i\rangle,\langle q,i+1\rangle\rangle\in E_{w}^{\mathcal{A}}$
iff $\langle p,a(i),q\rangle\in\Delta$. For a subset $T$ of $S$,
we say that a vertex $\langle p,i\rangle$ is a $T$-vertex if $p\in T$.
By definition $p\overset{w}{\longrightarrow}q$ iff there is a path
(in the directed sense) in $\mathcal{G}_{w}^{\mathcal{A}}$ from $\langle p,0\rangle$
to $\langle q,length(w)\rangle$ and $p\overset{w}{\underset{T}{\longrightarrow}}q$
if furthermore the path visits some $T$-vertex.

Finally we define the \emph{state complexity}%
\footnote{In some literature, instead of merely counting the number of states,
sizes of transition relations etc.\ are also taken into account to
better measure the sizes of automata. Here we prefer state complexity
because it is a measure easier to study, and its lower bound results
usually imply lower bounds on {}``size'' complexity, if the automata
witnessing the lower bound are over a not too large alphabet.%
} functions. Assume that $\mathcal{T}$ is either $\NFW$ or some common
type of $\omega$-automata. Then for a $\mathcal{T}$ automaton $\mathcal{A}$,
$C_{\mathcal{T}}(\mathcal{A})\ $is defined as the minimum number
of states of a $\mathcal{T}$ automaton that complements $\mathcal{A}$,
i.e., accepts $\mathcal{L}^{C}(\mathcal{A})$. For $n\geq1$, $C_{\mathcal{T}}(n)$
is the maximum of $C_{\mathcal{T}}(\mathcal{A})$ over all $\mathcal{T}$
automata with $n$ states. If indices are defined for $\mathcal{T}$,
then $C_{\mathcal{T}}(n,k)$ is the maximum of $C_{\mathcal{T}}(\mathcal{A})$
over all $\mathcal{T}$ automata with $n$ states and index $k$.

\section{The Full Automata Technique}\label{sec:full}

In the recently emerging area of state complexity (see \cite{Yu04}
for a survey) or in the theory of $\omega$-automata, we often concern
proving theorems of such flavor:

\begin{thm}
\label{thm:NFW}\cite{Jir05} For each $n\geq1$, there exists an
$\NFW$ $\mathcal{A}_{n}$ with $n$ states over $\{ a,b\}$ such
that $C_{\NFW}(\mathcal{A}_{n})\geq2^{n}$. 
\end{thm}
In other words, we want to prove a lower bound for the state complexity
of a transformation ($\NFW$ complementation in this case, can be
determinization etc.), and furthermore, we hope that the automata
family witnessing the lower bound ($(\mathcal{A}_{n})_{n\geq1}$ in
this case) is over a fixed small alphabet. Such claims are usually
difficult to prove. The apparently easy Theorem \ref{thm:NFW} was
not proved until 2005 by a very technical proof in \cite{Jir05}%
\footnote{The result is actually slightly stronger in that his $\mathcal{A}_{n}$
has only one initial state. (In some literature $\NFW$s are not allowed
to have multiple initial states.)%
}, after the efforts in \cite{SS78,Bir93,HK02}. To understand the
difficulty involved, we first review the traditional approach people
attempt at such results:

\begin{description}
\item [{Step\ I}] Identify an automata family $(\mathcal{A}_{n})_{n\geq1}$
with each $\mathcal{A}_{n}$ having $n$ states.
\item [{Step\ II}] Prove that to transform each $\mathcal{A}_{n}$ needs
a large state blow-up. 
\end{description}
Almost every known lower bound was obtained in this way, including
Theorem \ref{thm:NFW} and the aforementioned Michel's lower bound.
In such an approach, Step I is well-known to be difficult. Identifying
the suitable family $(\mathcal{A}_{n})_{n\geq1}$ requires both ingenuity
and luck. Even worse, most automata families that people try are natural
ones with simple structures, while the ones witnessing the desired
lower bound could be highly unnatural and complex. Finding the right
family $(\mathcal{A}_{n})_{n\geq1}$ seems to be a major obstacle
towards lower bound results.

Now we introduce the notion of full automata to circumvent this obstacle.

\begin{defn}
Given state set $S$, initial state set $I$, and extra components
$*$, a \emph{full automaton} $\mathcal{A}=(\Sigma,S,I,\Delta,\ast)$
is an automaton with alphabet $\Sigma=\mathcal{P}owerset(S\times S)$
and transition relation $\Delta$ defined as: for all $p,q\in S\mbox{ and }a\in\Sigma$,
$\langle p,a,q\rangle\in\Delta$ iff $\langle p,q\rangle\in a$. 

By definition, the alphabet contains every binary relation over $S$,
and therefore is of a big size of $2^{|S|^{2}}$. Due to such rich
alphabets, every automaton has some embedding in a full automaton
with the same number of states. It is then not difficult to see that
transforming an automaton can be reduced to transforming a full automaton,
and full automata are the most difficult automata to transform. 

To be specific, if we consider $\NFW$ complementation, then:
\end{defn}
\begin{thm}
\label{thm:completeness}For all $n\geq1$, $C_{\NFW}(n)=C_{\NFW}(\mathcal{A})$
for some full $\NFW$ $\mathcal{A}$ with $n$ states. 
\end{thm}
The theorem follows from the following lemma.

\begin{lem}
\label{lem:reduction}If $\mathcal{A}_{1}$ is an $\NFW$ with $n$
states, then there is a full $\NFW$ $\mathcal{A}_{2}$ with $n$
states such that $C_{\NFW}(\mathcal{A}_{2})\geq C_{\NFW}(\mathcal{A}_{1})$.
\end{lem}
\begin{proof}
By definition of $C_{\NFW}$, it suffices to show that for some full
$\NFW$ $\mathcal{A}_{2}$ with $n$ states, if there is an $\NFW$
$\mathcal{CA}_{2}\ $that complements $\mathcal{A}_{2}$, then there
is an $\NFW$ $\mathcal{CA}_{1}$ complementing $\mathcal{A}_{1}$
with the same number of states as $\mathcal{CA}_{2}$.

Let $\mathcal{A}_{1}=(\Sigma_{1},S_{1},I_{1},\Delta_{1},F_{1})$,
and consider the full $\NFW$ $\mathcal{A}_{2}=(\Sigma_{2},S_{1},I_{1},\Delta_{2},F_{1})$
with respect to $S_{1},I_{1}\mbox{ and }F_{1}$. For each $a_{1}\in\Sigma_{1}$,
define letter $\Delta_{1}(a_{1})$ in $\Sigma_{2}=\mathcal{P}(S_{1}\times S_{1})$
as: $\langle p_{1},q_{1}\rangle\in\Delta_{1}(a_{1})$ iff $\langle p_{1},a_{1},q_{1}\rangle\in\Delta_{1}$,
for all $p_{1},q_{1}\in S_{1}$. By definition of full automata, $\langle p_{1},a_{2},q_{1}\rangle\in\Delta_{2}$
iff $\langle p_{1},q_{1}\rangle\in a_{2}$, for all $p_{1},q_{1}\in S_{1},a_{2}\in\Sigma_{2}$.
So we have $\langle p_{1},a_{1},q_{1}\rangle\in\Delta_{1}$ iff $\langle p_{1},\Delta_{1}(a_{1}),q_{1}\rangle\in\Delta_{2}$,
for all $a_{1}\in\Sigma_{1},p_{1},q_{1}\in S_{1}$. For an arbitrary
word $\alpha=a(0)a(1)\dots a(l-1)\in\Sigma_{1}^{*}$, consider word
$\alpha^{\prime}=\Delta_{1}(a(0))\Delta_{1}(a(1))\dots\Delta_{1}(a(l-1))\in\Sigma_{2}^{*}$.
Then every state sequence $\rho_{1}=\rho_{1}(0)\rho_{1}(1)\dots\rho_{1}(l)\in S_{1}^{*}$
is a run of $\mathcal{A}_{1}$ over $\alpha$ iff $\rho_{1}$ is a
run of $\mathcal{A}_{2}$ over $\alpha^{\prime}$. Since $\mathcal{A}_{1}$
and $\mathcal{A}_{2}$ share the same initial and final state sets,
$\rho_{1}$ is successful iff $\rho_{2}$ is successful. So $\alpha\in\mathcal{L}(\mathcal{A}_{1})$
iff $\alpha^{\prime}\in\mathcal{L}(\mathcal{A}_{2})$.

Let $\mathcal{CA}_{2}=(\Sigma_{2},S_{C},I_{C},\Delta_{C},F_{C})$
be an $\NFW$ that complements $\mathcal{L}(\mathcal{A}_{2})$. So
$\alpha^{\prime}\in\mathcal{L}(\mathcal{A}_{2})$ iff $\alpha^{\prime}\notin\mathcal{L}(\mathcal{CA}_{2})$.
Define $\mathcal{CA}_{1}$ to be the $\NFW$ $(\Sigma_{1},S_{C},I_{C},\Delta_{C}^{\prime},F_{C})$,
where $\Delta_{C}^{\prime}$ is defined as $\langle p_{2},a_{1},q_{2}\rangle\in\Delta_{C}^{\prime}$
iff $\langle p_{2},\Delta_{1}(a_{1}),q_{2}\rangle\in\Delta_{C}$,
for all $p_{2},q_{2}\in S_{C}\mbox{ and }a_{1}\in\Sigma_{1}.$ Similarly
every state sequence $\rho_{C}=\rho_{C}(0)\rho_{C}(1)\dots\rho_{C}(l)\in S_{C}^{*}$
is a successful run of $\mathcal{CA}_{2}$ over $\alpha^{\prime}$
iff $\rho_{C}$ is a successful run of $\mathcal{CA}_{1}$ over $\alpha$.
So $\alpha^{\prime}\in\mathcal{L}(\mathcal{CA}_{2})$ iff $\alpha\in\mathcal{L}(\mathcal{CA}_{1})$.

Now for every $\alpha\in\Sigma_{1}^{\ast}$, $\alpha\in\mathcal{L}(\mathcal{A}_{1})$
iff $\alpha\notin\mathcal{L}(\mathcal{CA}_{1})$. Therefore $\mathcal{CA}_{1}$
with the same number of states as $\mathcal{CA}_{2}$ complements
$\mathcal{A}_{1}$ as required. 
\end{proof}
Theorem \ref{thm:completeness} implies that to prove a lower bound
for $\NFW$ complementation (without taking the size of the alphabet
into account), we can simply set $(\mathcal{A}_{n})_{n\geq1}$ to
be some family of full $\NFW$s in Step I. Similarly, the same applies
to $\NBW$ complementation:

\begin{thm}
\label{thm:nbw_completeness}For all $n\geq1$, $C_{\NBW}(n)=C_{\NBW}(\mathcal{A})$
for some full $\NBW$ $\mathcal{A}$ with $n$ states. 
\end{thm}
Now we apply full automata to obtain a simple proof of Theorem \ref{thm:NFW}.

\begin{proof}
(of Theorem \ref{thm:NFW}) We first prove a $2^{n}$ lower bound
for $C_{\NFW}(n)$. For each $n\geq1$, let $\mathcal{FA}_{n}=(\Sigma_{n},S_{n},I_{n},\Delta_{n},F_{n})$
be the full $\NFW$ with $S_{n}=I_{n}=F_{n}=\{ s_{0},\dots,s_{n-1}\}$.
It suffices to prove that $C_{\NFW}(\mathcal{FA}_{n})\geq2^{n}$. 

\begin{figure}[h]
\subfigure[\label{fig:uv}$u_T v_T$]{\includegraphics[scale=1.5]{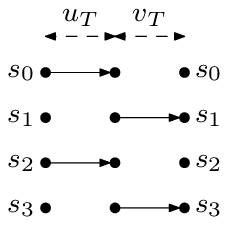}}\hspace{1.0in}\subfigure[$c_1 c_3\sim Id(T)$\label{fig:c1c3}]{\includegraphics[scale=1.5]{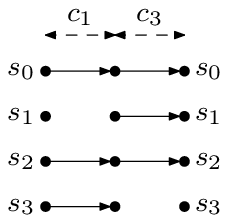}}

\subfigure[$abaaa\sim c_1$\label{fig:abaaa}]{\includegraphics[scale=1.5]{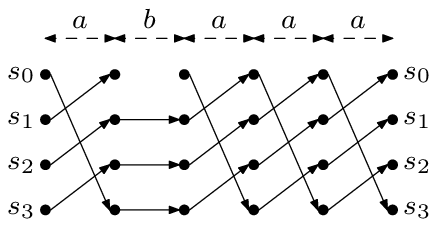}}\hspace{0.3in}\subfigure[$\mathcal{A}_4$\label{fig:a4}]{\includegraphics[scale=1.5]{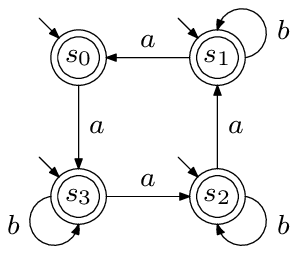}}

\caption{Examples}
\end{figure}

For each subset $T\subseteq S_{n}$, let $Id(T)$ denote the letter
$\{\langle q,q\rangle\mid q\in T\}$ and let $u_{T}=Id(T)$, $v_{T}=Id(S_{n}\backslash T)$.
Figure \ref{fig:uv} depicts one example of $u_{T}v_{T}$'s $\Delta$-graph.
Since all states in $\mathcal{FA}_{n}$ are both initial and final,
a word $w$ of length $l$ is accepted by $\mathcal{FA}_{n}$ iff
there is a path from an $\langle s_{i},0\rangle$ vertex to an $\langle s_{j},l\rangle$
vertex in the $\Delta$-graph of $w$ under $\mathcal{FA}_{n}$. In
particular $u_{T}v_{T}$ is not accepted by $\mathcal{FA}_{n}$. Suppose
that some $\NFW$ $\mathcal{CA}$ complements $\mathcal{FA}_{n}$.
So for each $T\subseteq S_{n}$, there is a state $\hat{q}_{T}$ of
$\mathcal{CA}$ such that $\hat{q}_{I}\overset{u_{T}}{\longrightarrow}\hat{q}_{T}$
and $\hat{q}_{T}\overset{v_{T}}{\longrightarrow}\hat{q}_{F}$ for
some initial state $\hat{q}_{I}$ and final state $\hat{q}_{F}$ of
$\mathcal{CA}$. If we prove that $\hat{q}_{T_{1}}\neq\hat{q}_{T_{2}}$
whenever $T_{1}\neq T_{2}$, then $\mathcal{CA}$ has at least $2^{n}$
states as required. Suppose by contradiction that $\hat{q}_{T_{1}}=\hat{q}_{T_{2}}$
for some $T_{1}\neq T_{2}$. W.l.o.g. there is a state $s$ of $\mathcal{FA}_{n}$
in $T_{1}\backslash T_{2}$. Then $s\overset{u_{T_{1}}}{\longrightarrow}s\overset{v_{T_{2}}}{\longrightarrow}s$
and hence $u_{T_{1}}v_{T_{2}}\in\mathcal{L}(\mathcal{FA}_{n})$. On
the other hand, for some initial state $\hat{q}_{I}$ and final state
$\hat{q}_{F}$ of $\mathcal{CA}$, $\hat{q}_{I}\overset{u_{T_{1}}}{\longrightarrow}\hat{q}_{T_{1}}=\hat{q}_{T_{2}}\overset{v_{T_{2}}}{\longrightarrow}\hat{q}_{F}$.
So $u_{T_{1}}v_{T_{2}}\in\mathcal{L}(\mathcal{CA})$, contradiction.

The above proof is not fully satisfying in that the automata family
witnessing the lower bound is over an exponentially growing alphabet.
To fix a binary alphabet and prove Theorem \ref{thm:NFW}, we introduce
a Step III in which we do {}``alphabet substitution'', as we now
illustrate.

We first refine the above proof of $C_{\NFW}(\mathcal{FA}_{n})\geq2^{n}$
by restricting the number of different letters involved. For two words
$u,v\in\Sigma_{n}^{\ast}$, we say that $u$ is \emph{equivalent}
to $v$ with respect to $\mathcal{FA}_{n}$, or simply $u\sim v$,
if for all $p,q\in S_{n}$, $p\overset{u}{\rightarrow}q$ iff $p\overset{v}{\rightarrow}q$.
A little thought shows that if we substitute each $Id(T)$ letter
used in the above proof by some equivalent words, the proof still
works. First we consider the alphabet $\{ c_{i}\}_{0\leq i<n}$ with
$c_{i}=Id(S_{n}\backslash\{ s_{i}\})$. Then for each $T\subseteq S_{n}$,
$Id(T)\sim\Pi_{s\notin T}c_{i}$, the concatenation of all $c_{i}$'s
with $s_{i}\notin T$ in lexicographical order (any other fixed order
will do). This is illustrated in Figure \ref{fig:c1c3}. Then consider
the alphabet $\{ a,b\}$ with $a=\{\langle s_{i+1},s_{i}\rangle\mid0\leq i<n-1\}\cup\{\langle s_{0},s_{n-1}\rangle\}$
and $b=Id(S_{n}\backslash\{ s_{0}\})$, then for each $0\leq i<n$,
$c_{i}\sim a^{i}ba^{n-i}$, as illustrated in Figure \ref{fig:abaaa}.
So if we substitute each letter $Id(T)$ in the above proof by the
equivalent word $\Pi_{s_{i}\notin T}a^{i}ba^{n-i}$, the proof still
works.

After the above refinement of the proof, the part of $\mathcal{FA}_{n}$
related to letters other than $\{ a,b\}$ is in fact irrelevant to
the proof. So $\mathcal{A}_{n}=\mathcal{FA}_{n}\upharpoonright\{ a,b\}$,
the restriction of $\mathcal{FA}_{n}$ to $\{ a,b\}$, or formally
the $\NFW$ $\mathcal{A}_{n}=(\{ a,b\}$, $S_{n}$, $I_{n}$, $\Delta_{n}\cap(S_{n}\times\{ a,b\}\times S_{n})$,
$F_{n})$, also satisfies that $C_{\NFW}(\mathcal{A}_{n})\geq2^{n}$,
as required ($\mathcal{A}_{4}$ is depicted in \ref{fig:a4}). 
\end{proof}
We call the above technique of setting $(\mathcal{A}_{n})_{n\geq1}$
to be a family of full automata and adding the step of alphabet substitution
the {}``full automata technique''. Setting $(\mathcal{A}_{n})_{n\geq1}$
to be full automata is crucial here, which in essence delays the trouble
of identifying $(\mathcal{A}_{n})_{n\geq1}$ to the later analysis
of transforming full automata. This makes our life easier because
the latter is usually playing with words, which is clearly easier
than constructing automata, especially with the rich alphabet of full
automata. As to the step of alphabet substitution, our experience
is that it could be technical some time, but rarely difficult.

\section{\buchi\ Complementation\label{sec:nbw}}

\subsection{Kupferman and Vardi's Construction}\label{sec:nbw:ranking}

We first briefly introduce the state-of-the-art construction for \buchi\ complementation
by Kupferman and Vardi in \cite{FKV06}, the idea of which is important
in our lower bound. Different from \cite{FKV06}, we will continue
to work with our $\Delta$-graphs rather than introducing the notion
of run graphs. For $x\in\mathbb{N}$, let $[x]$ denote the set $\{0,1,\dots,x\}$
and let $[x]^{odd}$ and $[x]^{even}$ denote the sets of odd and
even numbers in $[x]$ respectively.

\begin{defn}
Given an $\NBW$ $\mathcal{A}=(\Sigma,S,I,\Delta,F)$ of $n$ states,
and an $\omega$-word $\alpha$, a \emph{co-\buchi\ ranking} (C-Ranking
for short) for $\mathcal{G}_{\alpha}^{\mathcal{A}}$ (i.e.\ the $\Delta$-graph
of $\alpha$ under $\mathcal{A}$) is a partial function $f$ from
$V_{\alpha}^{\mathcal{A}}$ to the rank set $[2n-2]$ such that:
\begin{description}
\item [{(i)}] For all vertices $\langle q,l\rangle\in V_{\alpha}^{\mathcal{A}}$,
$f(\langle q,l\rangle)$ is undefined iff there is no path (in the
directed sense) from some $\langle q_{I},0\rangle$ vertex with $q_{I}\in I$
to $\langle q,l\rangle$.
\item [{(ii)}] For all vertices $\langle q,l\rangle\in V_{\alpha}^{\mathcal{A}}$,
if $f(\langle q,l\rangle)$ is odd, then $q\notin F$.
\item [{(iii)}] For all edges $\langle\langle q,l\rangle,\langle q^{\prime},l+1\rangle\rangle\in E_{\alpha}^{\mathcal{A}}$,
if $f(\langle q,l\rangle)$ is defined, then $f(\langle q,l\rangle)\geq f(\langle q^{\prime},l+1\rangle)$. 
\end{description}
We say that $f$ is \emph{odd} if for every path in $\mathcal{G}_{\alpha}^{\mathcal{A}}$,
there are infinitely many vertices that are assigned odd ranks by
$f$.
\end{defn}
\begin{lem}
\label{lem:ranking}\cite{KV01} The $\omega$-word $\alpha$ is not
accepted by $\mathcal{A}$ iff there is an odd C-ranking for $\mathcal{G}_{\alpha}^{\mathcal{A}}$. 
\end{lem}
\begin{proof}
We prove the \emph{if} direction here to give a sense of the idea
of C-ranking. For every infinite path from a $\langle q_{I},0\rangle$
vertex for some $q_{I}\in I$, the ranks along the path do not increase
by (iii) and so will get trapped in some fixed rank from some point
on. Since $f$ is odd, this fixed rank is odd, and thus by (ii), $F$-vertices
are never visited since then. In other words, every run of $\mathcal{A}$
over $\alpha$ visits $F$ finitely often and hence $\alpha$ is not
accepted by $\mathcal{A}$. 
\end{proof}
A \emph{level ranking}%
\footnote{Our definitions of level ranking and tight level ranking here are
slightly different from \cite{FKV06}.%
} for $\mathcal{A}$ is a partial function $g:S\longrightarrow[2n-2]$
such that if $g(q)$ is odd, then $q\notin F$. Each C-ranking can
be {}``sliced'' into such level rankings. It was shown in \cite{KV01}
that existence of an odd C-ranking for $\mathcal{G}_{\alpha}^{\mathcal{A}}$
can be decided by an $\NBW$ $\mathcal{CA}$ which guesses an odd
C-ranking level by level, and checks the validity in a local manner.
By Lemma \ref{lem:ranking}, $\mathcal{CA}$ complements $\mathcal{A}$.
In the construction of $\mathcal{CA}$, distinct sets of states are
used to handle different level rankings, and the number of such level
rankings is the major factor of the $(6n)^{n}$ blow-up.

We say that a level ranking $g$ for $\mathcal{A}$ is \emph{tight}
if (i): the maximum rank in the range of $g$ is some odd number $2m-1$
in $[2n-2]^{odd}$, and (ii): for every $j\in[2m]^{odd}$, there is
a state $q$ with $g(q)=j$. In such a case, $g$ is also called a
TL$(m)$-ranking (with $1\leq m<n$). It was further shown in \cite{FKV06}
that we can restrict attention to tight level rankings and use less
states in $\mathcal{CA}$. By a careful numerical analysis \cite{FKV06},
a $(0.97)^{n}$ upper bound was proved for the number of states of
$\mathcal{CA}$ and thus for \buchi\ complementation.

\subsection{Lower Bound}\label{sec:nbw:proof}

We turn now to lower bound. By Theorem \ref{thm:nbw_completeness},
it suffices to consider full $\NBW$s. We define $\mathcal{FB}_{n}$
for $n>1$ to be the full $\NBW$ $(\Sigma_{n},S_{n},I_{n},\Delta_{n},F_{n})$
with $I_{n}=\{ s_{0},\dots,s_{n-2}\}$, $F_{n}=\{ s_{f}\}$ and $S_{n}=I_{n}\cup F_{n}$.
We also use $S_{n}^{\prime}=I_{n}$ to denote the {}``main'' states.

We first try to construct an $\omega$-word $\alpha_{n}$ not accepted
by $\mathcal{FB}_{n}$ such that a great number of tight level rankings
would have to be present in every C-ranking for $\mathcal{G}_{\alpha_{n}}^{\mathcal{FB}_{n}}$.
Since the number of tight level rankings is the major factor of the
state blow-up in Kupferman and Vardi's construction, this would produce
a hard case for the construction. For such purpose, we consider a
special class of tight level rankings for $\mathcal{FB}_{n}$, $Q$-rankings.
We say that a TL$(m)$-ranking $g$ for $\mathcal{FB}_{n}$ is a $Q(m)$\emph{-ranking}
if $g\left(q\right)$ is defined for each $q\in S_{n}^{\prime}$ and
is undefined for $q=s_{f}$. We start defining our difficult $\omega$-word
$\alpha_{n}$ by defining its composing segments.

\begin{lem}
\label{lem:nbw:existence}For every pair of $Q$-rankings $(f,g)$,
there exists a word $w_{f,g}$ such that:
\begin{description}
\item [{(i)}] For all $p,q\in S_{n}^{\prime}$, $p\overset{w_{f,g}}{\longrightarrow}q$
iff ($f_{i}(p)>f_{i+1}(q)$ or $f_{i}(p)=f_{i+1}(q)\in[2m]^{odd}$).
\item [{(ii)}] For all $p,q\in S_{n}^{\prime}$, $p\overset{w_{f,g}}{\underset{F_{n}}{\longrightarrow}}q$
iff $f_{i}(p)>f_{i+1}(q)$.
\item [{(iii)}] For all $p,q\in S_{n}$, if $p\overset{w_{f,g}}{\longrightarrow}q$
then $p,q\notin F_{n}$. 
\end{description}
\end{lem}
\begin{proof}
We first illustrate the construction using a typical example depicted
in Fig. \ref{fig:wfg}. As in Fig. \ref{fig:wfg}, the vertices of
the $\Delta$-graph of $w_{f,g}$ are separated by the wider space
below $c(f,g)$ into two parts. We say that each $(s_{i},j)$ vertex
in the left part is ranked $f(s_{i})$ by $f$, and each $(s_{i},j)$
vertex in the right part is ranked $g(s_{i})$ by $g$. So when one
follows a path from a leftmost vertex $v_{1}$ to a rightmost vertex
$v_{2}$, either one goes to a next vertex with the same rank, or
one visits a $\langle s_{f},j\rangle$ vertex and then goes to a vertex
with a rank lower by one. This explains the \emph{only if} direction
of (ii). Also note that $v_{1}$ and $v_{2}$ cannot have the same
even ranks because in the middle of this process, one has to go to
a vertex with an odd rank to pass $c(f,g)$. So the \emph{only if}
direction in (i) holds too. For the \emph{if} directions of (i) and
(ii), suppose one wants to go from a leftmost vertex $v_{1}$ with
rank $r$ to a rightmost vertex $v_{2}$ with rank $r'$ and that
either $r>r'$ or $r=r'\in[2m]^{odd}$. Let $t$ be an odd rank such
that $r\geq t\geq r'$. Then by the construction, one can go from
$v_{1}$ to some vertex with rank $t$ in the left part, pass through
$c(f,g)$ with rank $t$, and then continue to go to $v_{2}$ in the
right part. Note that in the process, if rank ever decreases, then
an $\langle s_{f},j\rangle$ vertex must have been visited. So the
if directions of (i) and (ii) hold as well. Condition (iii) is obviously
true.

\begin{figure}[h]
\includegraphics[width=1\textwidth]{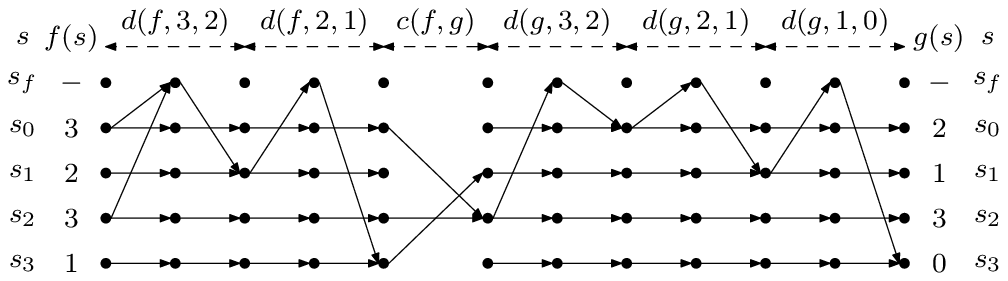}

\caption{\label{fig:wfg}$\Delta$-graph of $w_{f,g}$}
\end{figure}

For later purposes, we explicitly present our construction for $w_{f,g}$.
For a $Q(m)$-ranking $h$, we define the state sets $Rank_{h}(r)=\{ q\in S_{n}^{\prime}\mid r=h(q)\}$
for $r\in[2m]$ and $Odd_{h}$ to be the union of $Rank_{h}(r)$'s
with $r\in[2m]^{odd}$. Also for each $T\subseteq S_{n}^{\prime}$,
define letters in $\Sigma_{n}$ as $Id(T)=\{\langle q,q\rangle\mid q\in T\}$,
$TtoF(T)=Id(S_{n}^{\prime})\cup\{\langle q,s_{f}\rangle\mid q\in T\}$,
$FtoT(T)=Id(S_{n}^{\prime})\cup\{\langle s_{f},q\rangle\mid q\in T\}$
and $c(f,g)=\{\langle p,q\rangle\mid f(p)=g(q)\in[2m]^{odd}\text{, }p,q\in S_{n}^{\prime}\}$.
For a $Q(m)$-ranking $h$ and $r,r^{\prime}\in[2m]$, we write $d(h,r,r^{\prime})$
to denote the word $TtoF(Rank_{h}(r))\cdot FtoT(Rank_{h}(r^{\prime}))$.
Then if $r_{1},r_{2}\dots,r_{k}$ are the ranks in $[2m]$ that are
images of $h$ in descending order, we let $u_{h}=d(h,r_{1},r_{2})\cdot d(h,r_{2},r_{3})\cdot\dots\cdot d(h,r_{k-1},r_{k})$.
Finally, $w_{f,g}$ is defined to be $u_{f}\cdot c(f,g)\cdot u_{g}$.
\end{proof}
\begin{lem}
\label{lem:concatenation}Let $f_{0},f_{1},\dots,f_{l}$ be a list
of $Q(m)$-rankings with $l>0$, and let $w$ be the word $w_{f_{0},f_{1}}w_{f_{1},f_{2}}\dots w_{f_{l-1}f_{l}}$.
Also let $p,q\in S_{n}^{\prime}$, then:

(i) If $f_{0}(p)>f_{l}(q)$ or $f_{0}(p)=f_{l}(q)\in[2m]^{odd}$,
then $p\overset{w}{\longrightarrow}q$.

(ii) If $f_{0}(p)>f_{l}(q)$, then $p\overset{w}{\underset{F_{n}}{\longrightarrow}}q$.
\end{lem}
\begin{proof}
If $l=1$, then $w=w_{f_{0},f_{1}}$, and the properties follow from
Theorem \ref{lem:nbw:existence} trivially. So we assume that $l>1$.
Let $t$ be an odd rank such that $f_{0}(p)\geq t\geq f_{l}(q)$.
By definition of $Q(m)$-ranking, there exists a state sequence $q_{1},q_{2},\dots,q_{l-1}$
such that $f_{i}(q_{i})=t$ for all $1\leq i\leq l-1$ . So $q_{i}\overset{w_{f_{i},f_{i+1}}}{\underset{}{\longrightarrow}}q_{i+1}$
for all $1\leq i<l-1$. Also because $f_{0}(p)\geq t\geq f_{l}(q)$,
we have $p\overset{w_{f_{0},f_{1}}}{\underset{}{\longrightarrow}}q_{1}$
and $q_{l-1}\overset{w_{f_{l-1},f_{l}}}{\underset{}{\longrightarrow}}q$.
Concatenate these together, we have $p\overset{w}{\underset{}{\longrightarrow}}q$,
and (i) is satisfied. If $f_{0}(p)>f_{l}(q)$, then either $f_{0}(p)>t$
or $t>f_{l}(q)$, and hence either $p\overset{w_{f_{0},f_{1}}}{\underset{F_{n}}{\longrightarrow}}q_{1}$
or $q_{l-1}\overset{w_{f_{l-1},f_{l}}}{\underset{F_{n}}{\longrightarrow}}q$.
So $p\overset{w}{\underset{F_{n}}{\longrightarrow}}q$, and (ii) is
satisfied.
\end{proof}
Let $L(n,m)$ be the number of different $Q(m)$-rankings and let
$L(n)$ be $\underset{1\leq m<n}{\max}L(n,m)$. From now on we fix
$m$ such that $L(n)=L(n,m)$ and may simply write $L$ for $L(n)$.
Clearly there exists an infinite looping enumeration $f_{0},f_{1},\dots$
of $Q(m)$-rankings such that $f_{i}\neq f_{j}$ for all $i\neq j,0\leq i,j<L$,
and $f_{i}=f_{jL+i}$ for all $i,j\geq0$. Our {}``difficult'' $\omega$-word
$\alpha_{n}$ is then the $\omega$-word $w_{0}w_{1}\dots$ where
$w_{i}=w_{f_{i},f_{i+1}}$ for all $i\geq0$.

\begin{lem}
\label{lem:nbw:alpha_n}The $\omega$-word $\alpha_{n}$ is not in
$\mathcal{L}(\mathcal{FB}_{n})$. 
\end{lem}
\begin{proof}
If there is a successful run $\rho$ of $\mathcal{FB}_{n}$ over $\alpha_{n}$,
then there is an infinite state sequence $q_{0}q_{1}\dots\in S_{n}^{\omega}$
such that $q_{i}\overset{w_{i}}{\underset{}{\longrightarrow}}q_{i+1}$
for all $i\geq0$ and $q_{i}\overset{w_{i}}{\underset{F_{n}}{\longrightarrow}}q_{i+1}$
for infinitely many $i\in\mathbb{N}$. So by the construction of $w_{i}=w_{f_{i},f_{i+1}}$,
$f_{i}(q_{i})\geq f_{i+1}(q_{i+1})$ for all $i\geq0$ and $f_{i}(q_{i})>f_{i+1}(q_{i+1})$
for infinitely many $i\in\mathbb{N}$. This is impossible since $f_{0}(q_{0})$
is finite. 
\end{proof}
Recall that Kupferman and Vardi's construction uses distinct state
sets to handle different TL$(m)$-rankings. It turns out that if a
complement automaton of $\mathcal{FB}_{n}$ does not have as many
states as $Q(m)$-rankings, it would be {}``confused'' by $\alpha_{n}$
together with another complex $\omega$-word $\alpha'$ derived from
$\alpha_{n}$.

\begin{lem}
\label{lem:nbw:confusion}For each $n>1\ $and each $\omega$-automaton
$\mathcal{CA}$ with less than $L$ states, if $\rho$ is a run of
$\mathcal{CA}$ over $\alpha_{n}\notin\mathcal{L}(\mathcal{FB}_{n})$,
then there is a run $\rho^{\prime}$ of $\mathcal{CA}$ over some
$\omega$-word $\alpha^{\prime}\in\mathcal{L}(\mathcal{FB}_{n})$
with $Occ(\rho^{\prime})=Occ(\rho)$ and $Inf(\rho^{\prime})=Inf(\rho)$. 
\end{lem}
\begin{proof}
Suppose that $\mathcal{CA}=(\Sigma_{n},\hat{S},\hat{I},\hat{\Delta},Acc)$
is an $\omega$-automaton with less than $L$ states and $\rho=\rho(0)\rho(1)\dots\in\hat{S}^{\omega}$
is a run of $\mathcal{CA}$ over $\alpha_{n}$. Let $k_{0},k_{1},\dots$
be a number sequence such that $k_{0}=0$, $k_{i+1}-k_{i}=length(w_{i})$
for all $i\geq0$. So the $k_{i}$'s mark the positions where the
$w_{i}$'s concatenate. Therefore $\rho(k_{i})\overset{w_{i}}{\underset{}{\longrightarrow}}\rho(k_{i+1})$
for all $i\geq0$. Define for each $0\leq i<L$ the nonempty set:
\[
\hat{Q}_{i}=\{\hat{q}\in\hat{S}\mid\rho(k_{jL+i})=\hat{q}\text{ for infinitely many }j\in\mathbb{N}\}.\]
Since $\mathcal{CA}$ has less than $L$ states, there exists some
state $\hat{q}$ in $\hat{Q}_{i}\cap\hat{Q}_{j}$ for some $i\neq j,0\leq i,j<L$.
In particular one has, by definition, $f_{i}\neq f_{j}$. W.l.o.g.
there is a $q\in S_{n}^{\prime}$ with $f_{i}(q)>f_{j}(q)$. By definitions
of $\hat{Q}_{i}$ and $Occ(\rho)$, there is a $t_{1}\in\mathbb{N}$
sufficiently large such that $\rho(k_{t_{1}L+i})=\hat{q}$, every
state in $Occ(\rho)$ occurs in $\rho[0,k_{t_{1}L+i}]$, and that
$\rho(t^{\prime})\in Inf(\rho)$ for all $t^{\prime}>k_{t_{1}L+i}$.
By definitions of $Inf(\rho)$ and $\hat{Q}_{j}$, there is a sufficiently
large $t_{2}>t_{1}$ such that $\rho(k_{t_{2}L+j})=\hat{q}$ and every
state in $Inf(\rho)$ occurs in $\rho[k_{t_{1}L+i},k_{t_{2}L+j}]$.
Let $u=w_{0}\dots w_{t_{1}L+i-1}$ and $v=w_{t_{1}L+i}\dots w_{t_{2}L+j-1}$.
Finally let $\alpha^{\prime}$ be $uv^{\omega}$.

Let $q_{I}\in S_{n}^{\prime}$ be such that $f_{0}(q_{I})=2m-1\geq f_{i}(q)=f_{t_{1}L+i}(q)$.
By Lemma \ref{lem:concatenation}, $q_{I}\overset{u}{\underset{}{\longrightarrow}}q$.
Similarly, since $f_{t_{1}L+i}(q)=f_{i}(q)>f_{j}(q)=f_{t_{2}L+j}(q)$,
by Lemma \ref{lem:concatenation} we have $q\overset{v}{\underset{F_{n}}{\longrightarrow}}q$.
Together we have $q_{I}\overset{u}{\underset{}{\longrightarrow}}q\overset{v}{\underset{F_{n}}{\longrightarrow}}q\overset{v}{\underset{F_{n}}{\longrightarrow}}q\dots$
and $\alpha^{\prime}$ is accepted by $\mathcal{FB}_{n}$.

Finally, note that $\rho^{\prime}=\rho[0,k_{t_{1}L+i}]\cdot(\rho[k_{t_{1}L+i}+1,k_{t_{2}L+j}])^{\omega}$
is a run over $\alpha^{\prime}$, and we have guaranteed that $Occ(\rho^{\prime})=Occ(\rho)$
and $Inf(\rho^{\prime})=Inf(\rho)$ as required$.$ 
\end{proof}
\begin{thm}
\label{thm:nbw:buchi}For every $n>1$, $L(n)\leq C_{\NBW}(\mathcal{FB}_{n})\leq C_{\NBW}(n)$,
where $L(n)=\Theta((0.76n)^{n})$.
\end{thm}
\begin{proof}
By Lemma \ref{lem:nbw:confusion}, every $\NBW$ that complements
$\mathcal{FB}_{n}$ must have at least $L(n)$ states, otherwise both
$\alpha_{n}$ and $\alpha_{n}'$ would be accepted by $\mathcal{FB}_{n}$,
leading to contradiction. By a numerical analysis of $L(n)$ very
similar to the one in \cite{FKV06}, we have that $L(n)=\Theta((0.76n)^{n})$.
For completeness, we present the detail of the analysis in appendix.
\end{proof}

\subsection{Alphabet}\label{sec:nbw:alphabet}

Following the proof of Theorem \ref{thm:nbw:buchi}, one constructs
full $\NBW$s witnessing the lower bound over a very large alphabet,
which we rarely consider in practice. In this subsection, we show
that by using alphabet substitutions like in the proof of Theorem
\ref{thm:NFW}, the $\NBW$s witnessing the lower bound can be also
over a fixed alphabet.

We say two words $u$ and $v$ from $\Sigma_{n}^{\ast}$ are equivalent
with respect to $\mathcal{FB}_{n}$, or simply $u\approx v$, if for
all $p,q\in S_{n}^{\prime}$: (i) $p\overset{u}{\longrightarrow}q$
iff $p\overset{v}{\longrightarrow}q$, and, (ii) $p\overset{u}{\underset{F_{n}}{\longrightarrow}}q$
iff $p\overset{v}{\underset{F_{n}}{\longrightarrow}}q$. Then if one
replaces each letter involved in the lower bound proof by an equivalent
word over some alphabet $\Gamma$, one shows that $\mathcal{FB}_{n}\restriction\Gamma$
also witnesses the same $L(n)$ lower bound.

\begin{lem}
\label{lem:nbw:existence7}There is an alphabet $\Gamma$ of size
$7$ such that for each pair $\langle f,g\rangle$ of $Q(m)$-rankings
for $\mathcal{FB}_{n}$, there is a word in $\Gamma^{\ast}$ equivalent
to $w_{f,g}$.
\end{lem}
\begin{proof}
Let $\Gamma$ be the alphabet containing the following $7$ letters:
\begin{itemize}
\item $rotate=\{\langle s_{i+1},s_{i}\rangle\mid0\leq i<n-2\}\cup\{\langle s_{0},s_{n-2}\rangle,\langle s_{f},s_{f}\rangle\}$,
\item $clear0=Id(S_{n}\backslash\{ s_{0}\})$,
\item $swap01=(Id(S_{n}^{\prime})\cup\{\langle s_{0},s_{1}\rangle,\langle s_{1},s_{0}\rangle\})\backslash\{\langle s_{0},s_{0}\rangle,\langle s_{1},s_{1}\rangle\}$,
\item $copy01=Id(S_{n}^{\prime})\cup\{\langle s_{1},s_{0}\rangle\}$,
\item $0toF=Id(S_{n})\cup\{\langle s_{0},s_{f}\rangle\}$,
\item $Fto0=Id(S_{n})\cup\{\langle s_{f},s_{0}\rangle\}$,
\item $clearF=Id(S_{n}^{\prime})$.
\end{itemize}
Only three types of letters are relevant in the proof of Theorem \ref{thm:nbw:buchi}:
$TtoF(T)$, $FtoT(T)$ and $c(f,g)$. For each $T\subseteq S_{n}'$,
one can verify that:
\begin{itemize}
\item $TtoF(T)\approx clearF\cdot\dprod\limits _{s_{i}\in T}(rotate^{i}\cdot0toF\cdot rotate^{n-1-i})$.
\item $FtoT(T)\approx\dprod\limits _{s_{i}\in T}(rotate^{i}\cdot Fto0\cdot rotate^{n-1-i})\cdot clearF$. 
\end{itemize}
As to $c(f,g)$, the task is a bit more complicated, and let us view
it in a different way. For a word $w$, define set $r_{j}=\{ i|s_{i}\overset{w}{\longrightarrow}s_{j},0\leq i<n-1\}$
for every $0\leq j<n-1$. Clearly for two words $u\mbox{ and }v$,
the following are equivalent:
\begin{itemize}
\item $p\overset{u}{\longrightarrow}q$ iff $p\overset{v}{\longrightarrow}q$
for all $p,q\in S_{n}'$.
\item $r_{j}(u)=r_{j}(v)$ for all $0\leq j<n-1$.
\end{itemize}
So it is sufficient to find for each $c(f,g)$ a word $w$ over $\{ rotate,clear0,swap01,copy01\}$
such that $r_{j}(w)=r_{j}(c(f,g))$ for all $0\leq j<n-1$.

Appending each letter $a$ to the end of a word $w$ changes the content
of the $r_{i}(w)$'s. Consider these three types of words in $\Gamma_{}^{\ast}$:
\begin{enumerate}
\item $swap_{i,j}=\left\{ \begin{array}{cc}
rotate^{i}\cdot swap01\cdot rotate^{n-1-i} & \text{if }i+1=j\\
\begin{array}{c}
(swap_{i,i+1}\cdot swap_{i+1,i+2}\cdot\dots\cdot swap_{j-1,j})\\
\cdot(swap_{j-2,j-1}\cdot swap_{j-3,j-2}\cdot\dots\cdot swap_{i,i+1})\end{array} & \text{if }i+1<j\\
swap_{j,i} & \text{if }i>j\\
\text{the empty word} & \text{if }i=j\end{array}\right..$
\item $copy_{i,j}=\left\{ \begin{array}{cc}
swap01\cdot copy01\cdot swap01 & \text{if }i=1\text{ and }j=0\\
swap_{0,i}\cdot swap_{1,j}\cdot copy01\cdot swap_{1,j}\cdot swap_{0,i} & \text{otherwise}\end{array}\right..$
\item $clear_{i}=swap_{0,i}\cdot clear0\cdot swap_{0,i}$ 
\end{enumerate}
One can verify that appending a $swap_{i,j}$ to $w$ exchanges the
content of $r_{i}(w)$ and $r_{j}(w)$, appending a $copy_{i,j}$
sets $r_{i}(w)$ to be $r_{i}(w)\cup r_{j}(w)$, and appending a $clear_{i}$
empties $r_{i}(w)$. Obviously these three operations allow one to
reach arbitrary $(r_{i}(w))_{0\leq i<n-1}$ configurations, including
$(r_{i}(c(f,g)))_{0\leq i<n-1}$, as needed.
\end{proof}
So $\mathcal{B}_{n}=\mathcal{FB}_{n}\upharpoonright\Gamma$, the restriction
of $\mathcal{FB}_{n}$ to the alphabet $\Gamma$, satisfies that $C_{\NBW}(\mathcal{B}_{n})\geq L(n)$,
and we have:

\begin{thm}
\label{thm:nbw:alphabet}For each $n>1$, there exists an $\NBW$
$\mathcal{B}_{n}$ with $n$ states over a seven letters alphabet
such that $L(n)\leq C_{\NBW}(\mathcal{B}_{n})$.
\end{thm}

\subsection{Other Transformations}\label{sec:nbw:other}

Surprisingly, our lower bound on \buchi\ complementation extends
to almost every complementation or determinization transformation
of nondeterministic $\omega$-automata, via a reduction making use
of Lemma \ref{lem:nbw:confusion}.

\begin{thm}
\label{thm:nbw:main}For each $n>1$ and each common type $\mathcal{T}_{1}$
of nondeterministic $\omega$-automata, there exists a $\mathcal{T}_{1}$
automaton $\mathcal{A}_{n}$ with $n$ states over a fixed alphabet
such that:
\begin{description}
\item [{(i)}] For each common type $\mathcal{T}_{2}$, every $\mathcal{T}_{2}$
automaton that complements $\mathcal{L}(\mathcal{A}_{n})$ has at
least $L(n)$ states.
\item [{(ii)}] For each common type $\mathcal{T}_{2}$ that is not \buchi\ nor
generalized \buchi%
\footnote{Deterministic \buchi\ or generalized \buchi\ automata are strictly
weaker in expressive power than the other common types of $\omega$-automata.%
}, every deterministic $\mathcal{T}_{2}$ automaton that accepts $\mathcal{L}(\mathcal{A}_{n})$
has at least $L(n)$ states. 
\end{description}
\end{thm}
\begin{proof}
For each common type $\mathcal{T}_{1}$, by Fact \ref{fac:Loding},
there is a $\mathcal{T}_{1}$ automaton $\mathcal{A}_{n}$ equivalent
to $\NBW$ $\mathcal{FB}_{n}$ with also $n$ states \cite{Lod99}.
(i) Suppose that an automaton $\mathcal{CA}$ of a common type accepts
$\mathcal{L}^{C}(\mathcal{A}_{n})\mathcal{=L}^{C}(\mathcal{FB}_{n})$.
Since acceptance of $\omega$-automata of a common type only depends
on the $Inf$ set of a run, the claim can be obtained by applying
Lemma \ref{lem:nbw:confusion}. (ii) If some deterministic $\mathcal{T}_{2}$
automaton with less than $L(n)$ states accepts $\mathcal{L}(\mathcal{A}_{n})$,
and $\mathcal{T}_{2}$ is not \buchi\ or generalized \buchi, then
by Fact \ref{fac:Loding} there is a deterministic $\omega$-automaton
of a common type (not necessarily $\mathcal{T}_{2}$) complementing
$\mathcal{L}(\mathcal{A}_{n})$ with also less than $L(n)$ states
\cite{Lod99}, contrary to (i). Finally, the alphabet of $\mathcal{A}_{n}$
can be fixed like in the proof of Theorem \ref{thm:nbw:alphabet}. 
\end{proof}
For the transformations involved in this theorem, less than half already
had nontrivial lower bounds like $n!$ by Michel's proof or the bunch
of proofs by L\"oding \cite{Lod99}, while the others only have trivial
or weak $2^{\Omega(n)}$ lower bounds. These bounds are summarized
in Section \ref{sec:summary}.

\section{Complementation of Generalized \buchi\ Automata}\label{sec:ngbw}

We turn now to $\NGBW$ complementation. For $\NGBW$s, state complexity
is preferably measured in terms of both the number of states $n$
and index $k$, where index measures the size of the acceptance condition.
By applying full automata, doing a hard case analysis for the construction
in \cite{KV05} based on GC-ranking, and using a generalization of
Michel's technique, we prove an $(\Omega(nk))^{n}$ lower bound, matching
with the $(O(nk))^{n}$ bound in \cite{KV05}. This lower bound also
extends to the complementation of Streett automata and the determinization
of generalized \buchi\ automata into Rabin automata.

\subsection{Standard Full Generalized \buchi\ Automata \texorpdfstring{$\mathcal{FB}_{n,k}$}{FB{n,k}}}\label{sec:ngbw:fb_nk}

We first define full $\NGBW$ automata which we will show to witness
our desired lower bound.

We say a generalized \buchi\ acceptance condition $Acc=\{
F_{1},F_{2},\dots,F_{k}\}$ is \emph{minimal}, if no $F_{i},F_{j}$ pair
with $i\neq j$ satisfies that $F_{i}\subseteq F_{j}$. Note that if
such a pair exists, $F_{j}$ can be removed from $Acc$ without altering
the $\omega$-language defined. So we will only consider minimal
acceptance conditions. By the Sperner's theorem in
combinatorics \cite{Lub66}, if $Acc$ is minimal, then
$k\leq\binom{n}{\lfloor n/2\rfloor}$.

\begin{defn}
For $n>1$ and $1<k\leq\binom{n-1}{\lfloor(n-1)/2\rfloor}$, the \emph{standard
full $\NGBW$} $\mathcal{FB}_{n,k}=(\Sigma_{n},S_{n},I_{n},\Delta_{n},Acc_{n,k})$
is an $\NGBW$ with $\vert S_{n}\vert=n$, $I_{n}=S_{n}$ and a minimal
acceptance condition $Acc_{n,k}$. Let $s_{nf}$ be one of its state.
We denote $S_{n}\backslash\{ s_{nf}\}$ as $S_{n}^{\prime}$. $Acc_{n,k}$
is defined as an arbitrary fixed set $\{ F_{1},F_{2},\dots,F_{k}\}\subseteq\mathcal{P}(S_{n}^{\prime})$
such that: (i) $\vert F_{i}\vert=\lfloor(n-1)/2\rfloor$ for each
$F_{i}\in Acc_{n,k}$. (ii) For each $q\in S_{n}^{\prime}$, the number
of $F_{i}$'s in $Acc_{n,k}$ that do not contain $q$ is at least
$\lfloor k/2\rfloor$. 
\end{defn}
We must show that there is really such a minimal $Acc_{n,k}$
satisfying (i) and (ii). First let $Acc_{n,k}$ be a collection of
arbitrary $k$ distinct subsets of $S_{n}^{\prime}$ of
$\lfloor(n-1)/2\rfloor$ states and thus (i) is satisfied. Define
$\chi_{q}$ for each $q\in S_{n}^{\prime}$ as the number of $F_{i}$'s
in $Acc_{n,k}$ that contain $q$. By double counting, $\sum\limits
_{q\in S_{n}^{\prime}}\chi_{q}=\sum\limits _{i=1}^{k}\vert
F_{i}\vert$.  So if $\vert\chi_{p}-\chi_{q}\vert\leq1$ for all $p,q\in
S_{n}^{\prime}$, then for all $q\in S_{n}^{\prime}$,
$\chi_{q}\leq\lceil\frac{k\lfloor(n-1)/2\rfloor}{n-1}\rceil\leq\lceil
k/2\rceil$ and (ii) is also satisfied. Suppose $\chi_{p}-\chi_{q}>1$
for some $p,q\in S_{n}^{\prime}$. A little thought shows that there is
an $F_{i}\in Acc_{n,k}$ such that $p\in F_{i}$ and $(F_{i}\backslash\{
p\})\cup\{ q\}\notin Acc_{n,k}$.  Replace $F_{i}$ in $Acc_{n,k}$ by
$(F_{i}\backslash\{ p\})\cup\{ q\}$ and we make
$\vert\chi_{p}-\chi_{q}\vert$ strictly smaller. Repeat this till
$\vert\chi_{p}-\chi_{q}\vert\leq1$ for all $p,q\in S_{n}^{\prime}$.
Then condition (ii) is also satisfied.

\subsection{A Generalization of Michel's Technique}\label{sec:ngbw:technique}

We generalize the technique used in Michel's proof for \buchi\ complementation
\cite{Mic88} so that a tighter analysis of $\NGBW$ complementation
becomes possible.

\begin{defn}
A \emph{generalized co-\buchi}\ \emph{segment} (GC-segment for short)
$w$ of an $\NGBW$ $\mathcal{B}$ is a word such that $w^{\omega}\notin\mathcal{L}(\mathcal{B})$.
Two GC-segments $w_{1},w_{2}$ of $\mathcal{B}$ \emph{conflict} if
all $\omega$-words in the form $w_{1}^{k_{0}}(w_{1}^{k_{1}}w_{2}^{k_{2}})^{\omega},k_{i}>0$
are in $\mathcal{L}(\mathcal{B})$. A set $W$ of GC-segments of $\mathcal{B}$
is a \emph{conflict set} for $\mathcal{B}$ if every two distinct
GC-segments in $W$ conflict. 
\end{defn}
\begin{lem}
\label{lem:ngbw:technique}If $W$ is a conflict set for $\NGBW$
$\mathcal{B}$, then $C_{\NGBW}(\mathcal{B})\geq\vert W\vert$. 
\end{lem}
\begin{proof}
Suppose that some $\NGBW$ $\mathcal{CB}=(\Sigma,\hat{S},\hat{I},\hat{\Delta},\hat{F})$
complements $\mathcal{B}$, then for each GC-segment $w$ of $\mathcal{B}$
in $W$, $\mathcal{CB}$ accepts $w^{\omega}$. For every two distinct
GC-segments $w_{1},w_{2}\in W$, let $l_{1}=length(w_{1})$, $l_{2}=length(w_{2})$,
and let $\rho(0)\rho(1)\dots$ and $\rho^{\prime}(0)\rho^{\prime}(1)\dots$
be $\mathcal{CB}$'s two successful runs over $w_{1}^{\omega}$ and
$w_{2}^{\omega}$ respectively. Define \[
\hat{Q_{1}}=\{\hat{q}\in\hat{S}\mid\rho(i\cdot l_{1})=\hat{q}\mbox{ for infinitely many }i\in\mathbb{N\}}\]
 and \[
\hat{Q_{2}}=\{\hat{q}\in\hat{S}\mid\rho^{\prime}(i\cdot l_{2})=\hat{q}\mbox{ for infinitely many }i\in\mathbb{N\}.}\]
Clearly $\hat{Q_{1}}$ and $\hat{Q_{2}}$ are nonempty. It suffices
to show that $\hat{Q_{1}}\cap\hat{Q_{2}}=\emptyset,$ since it implies
that the number of states of $\mathcal{CB}$ is no less than the number
of GC-segments in $W$.

Suppose by contradiction that some $\hat{q}$ is in $\hat{Q_{1}}\cap\hat{Q_{2}}$.
By definition of $\hat{Q_{1}}$, there is a sufficiently large $k_{0}>0$
such that $\rho(k_{0}l_{1})=\hat{q}$ and for each $i\geq k_{0}l_{1}$,
$\rho(i)\in Inf(\rho)$. So $\rho[0,k_{0}l_{1}]$ is a finite run
over $w_{1}^{k_{0}}$ from some initial state $\hat{q}_{I}$ of $\mathcal{CB}$
to $\hat{q}$, i.e., $\hat{q}_{I}\overset{w_{1}^{k_{0}}}{\longrightarrow}\hat{q}$.
By definitions of $\hat{Q_{1}}$ and $Inf(\rho)$, there is a sufficiently
large $k_{1}>0$ such that $\rho((k_{0}+k_{1})l_{1})=\hat{q}$ and
in addition $\rho[k\cdot l_{1},(k_{0}+k_{1})l_{1}]$ is a finite run
from $\hat{q}$ to $\hat{q}$ over $w_{1}^{k_{1}}$ which visits every
state in $Inf(\rho)$. Similarly we have that for some $k_{0}^{\prime}$
and $k_{2}>0$, $\rho^{\prime}[k_{0}^{\prime}l_{2},(k_{0}^{\prime}+k_{2})l_{2}]$
is a finite run from $\hat{q}$ to $\hat{q}$ over $w_{2}^{k_{2}}$
which visits exactly every state in $Inf(\rho^{\prime})$. We construct
a new run as follows:\[
\rho_{new}=\rho[0,k_{0}l_{1}]\cdot\left(\rho[k_{0}l_{1}+1,(k_{0}+k_{1})l_{1}]\cdot\rho^{\prime}[k_{0}^{\prime}l_{2}+1,(k_{0}^{\prime}+k_{2})l_{2}]\right)^{\omega},\]
which is a run over $\alpha=w_{1}^{k_{0}}(w_{1}^{k_{1}}w_{2}^{k_{2}})^{\omega}$
with $Inf(\rho_{new})=Inf(\rho)\cup Inf(\rho^{\prime})$. As $\rho$
and $\rho^{\prime}$ are both successful, $\rho_{new}$ is also successful
by definition of generalized \buchi\ automata. So $\alpha$ is accepted
by $\mathcal{CB}$. However, as $w_{1}$ and $w_{2}$ conflict, $\alpha$
is accepted by $\mathcal{B}$ too, contradiction.
\end{proof}
\begin{cor}
\label{cor:ngbw:streett}If $W$ is a conflict set for $\NGBW$ $\mathcal{B}$,
then every $\NSW$ (nondeterministic Streett automaton) that complements
$\mathcal{B}$ has at least $\vert W\vert$ states. 
\end{cor}
\begin{proof}
Streett automata also satisfy that if $\rho$ and $\rho^{\prime}$
are both successful runs, then every run $\rho_{new}$ satisfying
$Inf(\rho_{new})=Inf(\rho)\cup Inf(\rho^{\prime})$ is also successful.
So the same proof as of Lemma \ref{lem:ngbw:technique} applies here.
\end{proof}

\subsection{A Conflict Set for
  \texorpdfstring{$\mathcal{FB}_{n,k}$}{FB{n,k}}}\label{sec:ngbw:conflict_set} 

It remains to define a large conflict set for $\mathcal{FB}_{n,k}$.
The following concept of pseudo generalized co-\buchi\ level ranking
is adapted from the concept of generalized co-\buchi\ level ranking
in the $\NGBW$ complementation construction in \cite{KV05}.

\begin{defn}
A \emph{pseudo generalized co-\buchi}\ \emph{level ranking} (PGCL-ranking
for short) \emph{}for $\mathcal{FB}_{n,k}$ is a pair $\langle f,g\rangle$
such that $f$ is a bijection from $S_{n}^{\prime}$ to $\{1,\dots,n-1\}$
and $g$ is a function from $S_{n}^{\prime}$ to $\{1,2,\dots,k\}$
such that each $q\in S_{n}^{\prime}$ is not contained in $F_{g(q)}$. 
\end{defn}
By definition of $\mathcal{FB}_{n,k}$, there are at least $\lfloor k/2\rfloor$
choices for the value of $g(q)$ for each $q\in S_{n}^{\prime}$.
So there are at least $(n-1)!\times(\lfloor k/2\rfloor)^{n-1}$ many
different PGCL-rankings, which is $(\Omega(nk))^{n}$ by Stirling's
formula.

Let $\mathcal{G}$ be a set of state sets. In the following, we use
notations in the form $p\overset{w}{\underset{\mathcal{G},!B}{\longrightarrow}}q$
to denote that there is a finite run over $w$ from $p$ to $q$ such
that the run visits every state set $F$ in $\mathcal{G}$, but it
does \emph{not} visit $B$. Either $\mathcal{G}$ or $B$ will be
omitted if is empty. In the following, we set \emph{$\mathcal{F}=\{ F_{1},\dots,F_{k}\}$.}

\begin{lem}
For each PGCL-ranking $\langle f,g\rangle$, there exists a word $seg_{f,g}$
with the properties that for all $p,q\in S_{n}^{\prime}:$
\begin{description}
\item [{(i)}] If $p=q$, i.e., $f(p)=f(q)$, then there is a unique finite
run of $\mathcal{FB}_{n,k}$ over $seg_{f,g}$ from $p$ to $q$,
and it is in the form $p\xrightarrow[\mathcal{F\backslash}F_{g(p)},!F_{g(p)}]{seg_{f,g}}q$.
\item [{(ii)}] If $f(p)>f(q)$, then there is a unique finite run of $\mathcal{FB}_{n,k}$
over $seg_{f,g}$ from $p$ to $q$, and it is in the form $p\overset{seg_{f,g}}{\underset{\mathcal{F}}{\longrightarrow}}q$.
\item [{(iii)}] If $f(p)<f(q)$, then there is no finite run of $\mathcal{FB}_{n,k}$
from $p$ to $q$ over $seg_{f,g}$.
\end{description}
\end{lem}
\begin{proof}
For notational convenience, we use notation like $\QATOPD[]{\oplus p_{1}\to p_{2},}{\ominus p_{3}\to p_{4},\ominus p_{5}\to p_{5}}$
to denote letter $\{\langle q,q\rangle\mid q\in S_{n}^{\prime}\}\cup\{\langle p_{1},p_{2}\rangle\}\backslash\{\langle p_{3},p_{4}\rangle,\langle p_{5},p_{5}\rangle\}$.
We also define a choice function $c(i,p)$ for each $i\in\{1,\dots,k\}$
and state $p\in S_{n}'$ with $g(p)\neq i$ such that $c(i,p)$ equals
to some arbitrary fixed element in $F_{i}\backslash F_{g(p)}$.

For each $r\in\{1,\dots,n-1\}$, let $p\in S_{n}^{\prime}$ be such
that $f(p)=r$, and define:\[
u_{r}=\prod\limits _{\substack{i\neq g(p),1\leq i\leq k\\
s=c(i,p)}
}\left[\begin{tabular}{ll}
 $\oplus p\rightarrow s$,  &  $\ominus p\rightarrow p$, \\
$\oplus s\rightarrow s_{nf}$,  &  $\ominus s\rightarrow s$\end{tabular}\right]\left[\begin{tabular}{ll}
 $\oplus s\rightarrow p$,  &  $\ominus p\rightarrow p$, \\
$\oplus s_{nf}\rightarrow s$,  &  $\ominus s\rightarrow s$\end{tabular}\right].\]
(Recall that $\Pi U$ means the concatenation of all words in $U$
in lexicographical order.) Then for each $q\in S_{n}'$, there is
a unique finite run over $u_{r}$ from $q$ to $q$, and it is in
the form $q\overset{u_{r}}{\underset{\mathcal{F\backslash}F_{g(p)},!F_{g(p)}}{\longrightarrow}}q$
if $p=q$, or $q\overset{u_{r}}{\underset{!F_{g(p)}}{\longrightarrow}q}$
otherwise.

For each $r=\{2,3,\dots,n-1\}$, let $p,q,s\in S_{n}^{\prime}$ be
such that $f(p)=r$, $f(q)=r-1$ and $s$ be an arbitrary state in
$F_{g(p)}$. Define:\[
v_{r}=\left[\begin{tabular}{ll}
 $\oplus p\rightarrow s$ ,  &  $\ominus s\rightarrow s$ ,\\
\multicolumn{2}{l}{$\oplus s\rightarrow s_{nf}$}\end{tabular}\right]\left[\begin{tabular}{ll}
 $\oplus s\rightarrow q$ ,  &  $\ominus s\rightarrow s$ ,\\
\multicolumn{2}{l}{$\oplus s_{nf}\rightarrow s$}\end{tabular}\right].\]
Then there is a unique finite run over $v_{r}$ from $p$ to $q$,
and it is in the form $p\overset{v_{r}}{\underset{F_{g(p)}}{\longrightarrow}}q$.
Also for every $q'\in S_{n}'$, there is a unique finite run over
$v_{r}$ from $q'$ to $q'$, and it is in the form $q'\overset{v_{r}}{\underset{!F_{g(p)}}{\longrightarrow}}q'$.

Finally let $seg_{f,g}$ be $u_{n-1}v_{n-1}u_{n-2}v_{n-2}\dots v_{2}u_{1}$.

To see that $seg_{f,g}$ satisfies the required properties, first
note that for all $p\in S_{n}^{\prime}$, $p\overset{u_{r}}{\underset{!F_{g(p)}}{\longrightarrow}}p$
and $p\overset{v_{r}}{\underset{!F_{g(p)}}{\longrightarrow}}p$. For
property (i), for every $p\in S_{n}^{\prime}$ with $f(p)=r$, there
exists a unique finite run over $seg_{f,g}$, and it is in the form:
\[
p\xrightarrow[!F_{g(p)}]{u_{n-1}v_{n-1}\dots u_{r+1}v_{r+1}}p\xrightarrow[\mathcal{F\backslash}F_{g(p)},!F_{g(p)}]{u_{r}}p\xrightarrow[!F_{g(p)}]{v_{r}u_{r-1}\dots v_{2}u_{1}}p,\]
that is, $p\xrightarrow[\mathcal{F\backslash}F_{g(p)},!F_{g(p)}]{seg_{f,g}}p$
as required. For property (ii), for every $p,q\in S_{n}^{\prime}$
with $f(p)=r_{1}>r_{2}=f(q)$, let $s_{r}\in S_{n}^{\prime}$ be such
that $f(s_{r})=r$ for each $r_{1}>r>r_{2}$. There is a unique finite
run over $seg_{f,g}$, and it is in the form: \begin{eqnarray*}
 &  & p\xrightarrow{u_{n-1}v_{n-1}\dots u_{r_{1}+1}v_{r_{1}+1}}p\xrightarrow[\mathcal{F\backslash}F_{g(p)},!F_{g(p)}]{u_{r_{1}}}p\xrightarrow[F_{g(p)}]{v_{r_{1}}}s_{r_{1}-1}\\
 &  & \xrightarrow{u_{r_{1}-1}v_{r_{1}-1}}s_{r_{1}-2}\dots s_{r_{2}+1}\xrightarrow{u_{r_{2}+1}v_{r_{2}+1}}q\xrightarrow{u_{r_{2}}\dots v_{2}u_{1}}q,\end{eqnarray*}
that is, $p\overset{seg_{f,g}}{\underset{\mathcal{F}}{\longrightarrow}}q$
as required. Property (iii) is easy to verify. 
\end{proof}
\begin{rem}
\label{rem:polysize}From the proof of the above lemma, it follows
that an alphabet of size polynomial in $n$ is sufficient to describe
$\{ seg_{f,g}|f,g\mbox{ are PGCL-rankings}\}$.
\end{rem}
\begin{lem}
For each PGCL-ranking $\langle f,g\rangle$ for $\mathcal{FB}_{n,k}$,
word $seg_{f,g}$ is a GC-segment of $\mathcal{FB}_{n,k}$. 
\end{lem}
\begin{proof}
Let $l=length(seg_{f,g})$, and let $\rho=\rho(0)\rho(1)\dots$ be
a run of $\mathcal{FB}_{n,k}$ over $seg_{f,g}^{\omega}$ in the form
$\rho(0)\overset{seg_{f,g}}{\longrightarrow}\rho(l)\overset{seg_{f,g}}{\longrightarrow}\rho(2l)\dots$.
Note that by the construction of $seg_{f,g}$, $\rho(i\cdot l)\in S_{n}^{\prime}$
and $f(\rho(i\cdot l))$ is defined for all $i\geq0$. Then by property
(iii), $f(\rho(0))\geq f(\rho(l))\geq f(\rho(2l))\geq\dots$ and then
for some $t\in\mathbb{N}$, $f(\rho(t^{\prime}\cdot l))=f(\rho(t\cdot l))$
for all $t^{\prime}>t$, that is $\rho(t^{\prime}\cdot l)=\rho(t\cdot l)$
for all $t^{\prime}>t$ since $f$ is a bijection. Let $j=g(\rho(t\cdot l))$.
By property (i), $F_{j}$ is not visited in $\rho[t'\cdot l,(t^{\prime}+1)\cdot l]$
for all $t^{\prime}\geq t$. So $Inf(\rho)\cap F_{j}=\emptyset$ and
hence $seg_{f,g}^{\omega}$ is not accepted by $\mathcal{FB}_{n,k}$. 
\end{proof}
\begin{lem}
\label{lem:ngbw:conflict_set}The set $W=\{ seg_{f,g}\mid\langle f,g\rangle\text{ is a PGCL-ranking for }\mathcal{FB}_{n,k}\}$
is a conflict set of size $(\Omega(nk))^{n}$ for $\mathcal{FB}_{n,k}$. 
\end{lem}
\begin{proof}
Suppose $\langle f_{1},g_{1}\rangle$ and $\langle f_{2},g_{2}\rangle$
are two distinct PGCL-rankings. Let $w_{1}=seg_{f_{1},g_{1}}$ and
$w_{2}=seg_{f_{2},g_{2}}$. There are two cases.
\begin{description}
\item [{Case}] I: $f_{1}$ and $f_{2}$ are two different bijections. So
there exist $p,q\in S_{n}^{\prime}$ such that $f_{1}(p)>f_{1}(q)$
and $f_{2}(p)<f_{2}(q)$. By property (i), $p\overset{w_{1}}{\underset{}{\longrightarrow}}p$,
$q\overset{w_{2}}{\underset{}{\longrightarrow}}q$ and so $p\overset{w_{1}^{m-1}}{\underset{}{\longrightarrow}}p,q\overset{w_{2}^{m-1}}{\underset{}{\longrightarrow}}q$
for all $m>0$. By property (ii), $p\overset{w_{1}}{\underset{\mathcal{F}}{\longrightarrow}}q$
and $q\overset{w_{2}}{\underset{\mathcal{F}}{\longrightarrow}}p$.
So for all $m>0$, $p\overset{w_{1}^{m}}{\underset{\mathcal{F}}{\longrightarrow}}q$
and $q\overset{w_{2}^{m}}{\underset{\mathcal{F}}{\longrightarrow}}p$.
Now for every $\omega$-word $\alpha$ in the form $w_{1}^{k_{0}}(w_{1}^{k_{1}}w_{2}^{k_{2}})^{\omega}$,
$k_{i}>0$, we construct a successful run over $\alpha$ as $p\overset{w_{1}^{k_{0}}}{\underset{}{\longrightarrow}}p\overset{w_{1}^{k_{1}}}{\underset{\mathcal{F}}{\longrightarrow}}q\overset{w_{2}^{k_{2}}}{\underset{\mathcal{F}}{\longrightarrow}}p\overset{w_{1}^{k_{1}}}{\underset{\mathcal{F}}{\longrightarrow}}q\overset{w_{2}^{k_{2}}}{\underset{\mathcal{F}}{\longrightarrow}}p\dots$.
So $\alpha$ is accepted by $\mathcal{FB}_{n,k}$ and $w_{1}$ conflicts
with $w_{2}$.
\item [{Case}] II: $f_{1}=f_{2}$ but $g_{1}\neq g_{2}$. Let $p\in S_{n}^{\prime}$
be such that $g_{1}(p)\neq g_{2}(p)$. By property (i), $p\xrightarrow[\mathcal{F\backslash}F_{g_{1}(p)},!F_{g_{1}(p)}]{w_{1}}p$
and $p\xrightarrow[\mathcal{F\backslash}F_{g_{2}(p)},!F_{g_{2}(p)}]{w_{2}}p$.
As $g_{1}(p)\neq g_{2}(p)$, $p\xrightarrow[\mathcal{F}]{w_{1}^{k_{1}}w_{2}^{k_{2}}}p$
for every $k_{1},k_{2}>0$. Now for every $\omega$-word $\alpha$
in the form $w_{1}^{k_{0}}(w_{1}^{k_{1}}w_{2}^{k_{2}})^{\omega}$,
$k_{i}>0$, we construct a successful run over $\alpha$ as $p\overset{w_{1}^{k_{0}}}{\underset{}{\longrightarrow}}p\xrightarrow[\mathcal{F}]{w_{1}^{k_{1}}w_{2}^{k_{2}}}p\xrightarrow[\mathcal{F}]{w_{1}^{k_{1}}w_{2}^{k_{2}}}p\dots$.
So $\alpha$ is accepted by $\mathcal{FB}_{n,k}$ and $w_{1}$ conflicts
with $w_{2}$. 
\end{description}
Finally, the size of $W$ is just the number of different PGCL-rankings
for $\mathcal{FB}_{n,k}$, which is $(\Omega(nk))^{n}$. 
\end{proof}

\subsection{Results}

\begin{thm}
\label{thm:ngbw:lowerbound}For $n>1$ and $1<k\leq$$\binom{n-1}{\lfloor(n-1)/2\rfloor}$,
$C_{\NGBW}(n,k)=(\Omega(nk))^{n}$. 
\end{thm}
\begin{proof}
The theorem follows from Lemma \ref{lem:ngbw:technique} and Lemma
\ref{lem:ngbw:conflict_set} directly. 
\end{proof}
This matches neatly%
\footnote{The gap hidden in the notation $(\Theta(nk))^{n}$ can be at most
$c^{n}$ for some $c$, while the gap hidden in the more widely used
notation $2^{\Theta(n\log nk)}$ can be as large as $(nk)^{n}$.%
} with the $(O(nk))^{n}$ construction in \cite{KV05}, and thus settles
the state complexity of $\NGBW$ complementation. Like Michel's result,
this lower bound can be extended to NSW complementation and the determinization
of $\NGBW$ into $\DRW$ (state complexity denoted by $D_{\NGBW\rightarrow\DRW}(n,k)$):

\begin{thm}
\label{thm:ngbw:consequence}For all $n>1$ and $1<k\leq$$\binom{n-1}{\lfloor(n-1)/2\rfloor}$,
$C_{\NSW}(n,k)=(\Omega(nk))^{n}$ and $D_{\NGBW\rightarrow\DRW}(n,k)=(\Omega(nk))^{n}$. 
\end{thm}
\begin{proof}
By Fact \ref{fac:Loding} there is an $\NSW$ $\mathcal{S}_{n,k}$
equivalent to each $\mathcal{FB}_{n,k}$ with the same number of states
and the same index. By Corollary \ref{cor:ngbw:streett} and Lemma
\ref{lem:ngbw:conflict_set}, every $\NSW$ that complements $\mathcal{FB}_{n,k}$
has $(\Omega(nk))^{n}$ states. So $C_{\NSW}(\mathcal{S}_{n,k})=(\Omega(nk))^{n}$
and $C_{\NSW}(n,k)=(\Omega(nk))^{n}$.

Suppose by contradiction that $\mathcal{R}$ is a $\DRW$ with less
than $|W|$ states that accepts $\mathcal{L}(\mathcal{FB}_{n,k})$,
then by Fact \ref{fac:Loding} there is a $\DSW$ $\mathcal{S}$ complementing
$\mathcal{FB}_{n,k}$ with the same number of states as $\mathcal{R}$,
contrary to Corollary \ref{cor:ngbw:streett}. So $D_{\NGBW\rightarrow\DRW}(n,k)=(\Omega(nk))^{n}$. 
\end{proof}
\begin{rem}
For the above lower bound, by Remark \ref{rem:polysize}, the alphabet
involved in the proof is of a size polynomial in $n$. It seems difficult
to fix a constant alphabet, but we conjecture this to be possible
if we aim at a weaker bound like $2^{\Omega(n\log nk)}$.
\end{rem}

\section{Summary}\label{sec:summary}

In the following table, we briefly summarize our lower bounds. Here
{}``Any'' means any common type of nondeterministic $\omega$-automata
(and the two Any's can be different). {}``co.'' means complementation
and {}``det.'' means determinization. {}``L.B.'' /{}``U.B.''
stands for lower/upper bound. Weak $2^{\Omega(n)}$ lower bounds are
considered trivial.

\vspace{1em}

\begin{center}
\begin{tabular}{|c|c|c|c|c|}
\hline 
\# &
Transformation &
Previous L.B. &
Our L.B. &
Known U.B. \tabularnewline
\hline 
1 &
$\NBW\overset{\text{co.}}{\longrightarrow}\NBW$ &
$\Omega((0.36n)^{n})$ \cite{Mic88} &
$\Omega((0.76n)^{n})$ &
$O((0.97n)^{n})$ \cite{FKV06} \tabularnewline
\hline 
2 &
Any$\overset{\text{co. or det.}}{\longrightarrow}\text{Any}$ &
trivial or $n!$ \cite{Lod99} &
$2^{\Omega(n\log n)}$ &
- \tabularnewline
\hline 
3 &
$\NBW$ $\overset{\text{det.}}{\longrightarrow}$ DMW &
trivial%
\footnote{But if size complexity is concerned, rather than state complexity,
then Safra proved that the transformation is inherently doubly exponential
\cite{Saf89}. %
}&
$2^{\Omega(n\log n)}$ &
$2^{O(n\log n)}$ \cite{Saf89} \tabularnewline
\hline 
4 &
$\text{NRW}\overset{\text{co.}}{\longrightarrow}\text{NRW}$ &
trivial%
\footnote{As pointed to us by Moshe Vardi, if size complexity is concerned,
then an $2^{\Omega(n\log n)}$ lower bound follows from Michel's lower
bound.%
} &
$2^{\Omega(n\log n)}$ &
$2^{O(nk\log n)}$ \cite{KV05a} \tabularnewline
\hline 
5 &
$\NGBW\overset{\text{co.}}{\longrightarrow}\NGBW$ &
$\Omega((n/e)^{n})$ \cite{Mic88} &
$(\Omega(nk))^{n}$ &
$(O(nk))^{n}$ \cite{KV05} \tabularnewline
\hline 
6 &
$\NSW\overset{\text{co.}}{\longrightarrow}\NSW$ &
$\Omega((n/e)^{n})$ \cite{Lod99} &
$(\Omega(nk))^{n}$ &
$2^{O(nk\log(nk))}$ \cite{KV05a} \tabularnewline
\hline 
7 &
$\NGBW\overset{\text{det.}}{\longrightarrow}\DRW$ &
$\Omega((n/e)^{n})$ \cite{Lod99} &
$(\Omega(nk))^{n}$ &
$2^{O(nk\log(nk))}$ \cite{Saf89} \tabularnewline
\hline
\end{tabular}
\par\end{center}

\vspace{1em}

In particular, lower bound \#2 implies that the $2^{\Omega(n\log n)}$
blow-up is inherent in the complementation and determinization of
nondeterministic $\omega$-automata, corresponding to the $2^{n}$
blow-up of finite automata. The special case \#3 justifies that Safra's
construction is optimal in state complexity for the determinization
of \buchi\ automata into Muller automata. We single out this result
because this determinization construction is touched in almost every
introductory material on $\omega$-automata, and its optimality problem
was explicitly left open in \cite{Lod99}.

For many of these transformations, it is still interesting to try
to narrow the complexity gap, and here we discuss three of them. First,
the complexity gap of \buchi\ complementation, although significantly
narrowed, is still exponential. By analyzing the difference between
the lower and upper bounds, one can find that the gap is mainly caused
by the use of the state component $O$ in \cite{FKV06} to maintain
the states along paths that have not visited an odd vertex since the
last time $O$ has been empty. So we should investigate how many states
are really necessary for such a purpose. Second, for Streett complementation,
the gap is still quite large. We feel that efforts should be first
taken to optimize the construction in \cite{KV05a}. Third, it is
interesting to see if an $\Omega(n^{n})$ or similar lower bound exists
for the determinization of $\NBW$s into Muller or Rabin automata.
Such would imply that determinization is harder than complementation
for $\omega$-automata, unlike the case of automata over finite words.
Of course, one can also work on the reverse direction, trying to design
ranking based constructions for determinization, which could have
good complexity bound as well as better applicability to practice.

Finally, we remark that the full automata technique has been quite
essential in obtaining our lower bound results. It is also possible
to extend the full automata technique to other kinds of automata,
like alternating automata or tree automata. We hope that the full
automata technique will stimulate the discovery of new results in
automata theory.

\begin{acknowledgement*}
I thank Orna Kupferman and Moshe Vardi for the insightful discussion
and the extremely valuable suggestions. I thank Enshao Shen for his
kind support and guidance. I also thank the anonymous referees for
the detailed and useful comments.
\end{acknowledgement*}
\bibliographystyle{alpha}
\bibliography{all4lmcs}

\appendix

\section{Numerical Analysis of \texorpdfstring{$L(n)$}{L(n)}}

In this section, we prove that $L(n)=\Theta((0.76n)^{n})$. The analysis
is very similar to the one in \cite{FKV06}, but we still present
it here for completeness. In the following, we write $f(n)\approx g(n)$
if two functions differ by only a polynomial factor in $n$. For example,
by Stirling's formula, $n!\approx(n/e)^{n}$. 

Let $T(n,m)$ denote the number of functions from $\left\{ 1\dots n\right\} $
onto $\left\{ 1\dots m\right\} $. The following estimate of $T(n,m)$
is implicit in Temme \cite{Tem93}:

\begin{lem}
\cite{Tem93}For $0<\beta<1$, let $x$ be the positive real number
solving $\beta x=1-e^{-x}$, and let $a=-\ln x+\beta\ln(e^{x}-1)-(1-\beta)+(1-\beta)\ln(1/\beta-1)$.
Then $T(n,\left\lfloor \beta n\right\rfloor )\approx\left(M\left[\beta\right]n\right)^{n}$,
where $M\left[\beta\right]=e^{a-\beta}\left(\frac{\beta}{1-\beta}\right)^{1-\beta}$.
\end{lem}
To prove a lower bound for $L(n)$, we first express $L(n,m)$ in
the following form:

\begin{lem}
$L(n,m)=\sum_{t=m}^{n-1}\binom{n-1}{t}T(t,m)m^{n-1-t}$ .
\end{lem}
\begin{proof}
To count the number of different $Q(m)$-ranking, we fix $t$, which
denotes the number of states that have odd ranks. Then there are $\binom{n-1}{t}$
ways to choose which $t$ states have odd ranks, and there are $T(t,m)$
ways to assign these $t$ states the $m$ different odd ranks. Moreover,
for each of the other $n-1-t$ states in $S_{n}'$, there are $m$
ways to choose which even rank it is assigned.
\end{proof}
\begin{thm}
$L(n)=\Omega(\left(c_{l}n\right)^{n})$, where $c_{l}=0.76$.
\end{thm}
\begin{proof}
By the previous lemma, $L(n)=\underset{m=1\dots n-1}{\max}\sum_{t=m}^{n-1}\binom{n-1}{t}T(t,m)m^{n-1-t}$.
Since we do not care about polynomial factors, $\sum_{t=m}^{n-1}$
can be replaced by $\underset{t=m\dots n-1}{\max}$, and we can replace
$m!$ by $(m/e)^{m}$ and $\binom{n-1}{t}$ by $\frac{n^{n}}{t^{t}(n-t)^{n-t}}$
as well. Also let $\gamma=m/n$ and $\beta=t/n$, then we have:
\end{proof}
$L(n)\approx\underset{0<\gamma\leq\beta<1}{\max}n^{n}(\beta n)^{-\beta n}((1-\beta)n)^{-(1-\beta)n}\cdot(M[\gamma/\beta]\beta n)^{\beta n}\cdot(\gamma n)^{n-1-\beta n}$

$\approx\underset{0<\gamma\leq\beta<1}{\max}(h(\beta,\gamma)n)^{n}$,
where $h\left(\beta,\gamma\right)=(1-\beta)^{\beta-1}(M[\gamma/\beta])^{\beta}\gamma^{1-\beta}$.

Computed by the Mathematica software, $h(\beta,\gamma)=0.7645$ when
$\beta=0.7236,\gamma=0.5744$. So $(0.76n)^{n}$ is an asymptotic
lower bound for $L(n)$.

\end{document}